\documentstyle[11pt]{article}
\begin{document}
\begin{titlepage}
\null
\begin{flushright}
DFTT-12/93 \\
March, 1993
\end{flushright}
\vspace{1cm}
\begin{center}
{\Large\bf
Condensation of Handles \\
in the Interface of 3D Ising Model}
\end{center}
\vskip 1.3cm
\centerline{ M. Caselle, F. Gliozzi and S. Vinti}
\vskip .6cm
\centerline{\sl  Dipartimento di Fisica
Teorica dell'Universit\`a di Torino}
\centerline{\sl Istituto Nazionale di Fisica Nucleare,Sezione di Torino}
\centerline{\sl via P.Giuria 1, I-10125 Turin,
Italy\footnote{{\it email address}:~~~~~~Decnet=
(31890::CASELLE,GLIOZZI,VINTI)\\
\centerline{Internet=CASELLE(GLIOZZI)(VINTI)@TORINO.INFN.IT}}}
\vskip 2.5cm
\vskip 1.5cm

\begin{abstract}
We analyze the microscopic, topological structure of the interface
between domains of opposite magnetization in 3D Ising model near the
critical point. This interface exhibits a fractal behaviour with a high
density of handles. The mean area is an almost linear function of the
genus. The entropy exponent  is affected by strong finite-size effects.
\end{abstract}

\normalsize
\end{titlepage}
\baselineskip=0.8cm
\setcounter{footnote}{0}

\newcommand{\eq}{\begin{equation}}
\newcommand{\en}{\end{equation}}
\newcommand{\eqa}{\begin{eqnarray}}
\newcommand{\ena}{\end{eqnarray}}
\newcommand{\spz}{\hspace{0.7cm}}
\newcommand{\lbl}{\label}
\newcommand{\lhi}{\hat\lambda_{i}}


\newcommand{\NP}[1]{Nucl.\ Phys.\ {\bf #1}}
\newcommand{\PL}[1]{Phys.\ Lett.\ {\bf #1}}
\newcommand{\NC}[1]{Nuovo Cim.\ {\bf #1}}
\newcommand{\CMP}[1]{Comm.\ Math.\ Phys.\ {\bf #1}}
\newcommand{\PR}[1]{Phys.\ Rev.\ {\bf #1}}
\newcommand{\PRL}[1]{Phys.\ Rev.\ Lett.\ {\bf #1}}
\newcommand{\MPL}[1]{Mod.\ Phys.\ Lett.\ {\bf #1}}
\newcommand{\IJMP}[1]{Int.\ J.\ Mod.\ Phys.\ {\bf #1}}
\newcommand{\JETP}[1]{Sov.\ Phys.\ JETP {\bf #1}}
\newcommand{\TMP}[1]{Teor.\ Mat.\ Fiz.\ {\bf #1}}

{\bf 1. Introduction}
\vskip .3cm
The interface in 3D statistical systems above the roughening
temperature behaves like a free, fluctuating surface with a  topology
determined by the boundary conditions. Recently, new computational
algorithms have been applied to these systems [1-6],  providing us with
a powerful tool for testing our ideas on the behaviour of fluctuating
surfaces. Alternatively, these surfaces can be thought of as space-time
histories of closed strings which can be used to describe the infrared
behaviour of the dual $ Z_2$ gauge theory.
\vskip .3 cm
In a previous paper \cite{cgv} we studied the free-energy
of the interface as a function of its shape. We found rather strong
 shape effects which are  accurately described by the gaussian
limit of the Nambu-Goto action.

This observation suggests that  interface configurations are mostly made
of smooth surfaces subjected to long-wavelength fluctuations which
account for the observed finite-size effects. On the other hand, this
picture seems to strongly disagree with the process of crumpling, which
should take place for random surfaces embedded in a target space of
dimension $d\geq1$. In other words, if the interface is described by the
Nambu-Goto action, it should take the shape of a branched polymer,
rather than that of a smooth surface.
\vskip .3 cm
In this letter we face  this  dilemma by studying
the interface from a microscopic point of view. Actually, it turns out
that the interface at small scales is much more similar to a sponge than
to a smooth surface. In particular there is a strong, almost linear
correlation between the area and the genus of the surface, indicating
the formation of a large number of microscopic handles which is
proportional to its area.
Yet, the partition function summed over all  the genera behaves like
that of a smooth toroidal surface: this means that the huge number of
microscopic handles, produced by the instability toward crumpling, have
as the net effect a simple non-perturbative renormalization of the
physical quantities associated to the surface, according to an old
conjecture on string theory \cite{sd}.
\vskip .3 cm
{\bf 2. The method}
\vskip .3cm
The first problem to be solved in order to study the microscopic
structure of the interface is to express the $3D$ Ising model or its
dual gauge version as a gas of suitable closed surfaces. There are
essentially two different ways to do it. One is based on the strong
coupling expansion of the dual gauge model \cite{ic} and involves also
self-intersecting and non-orientable surfaces. Here instead we identify
the surfaces as the Peierls interfaces of an arbitrary Ising spin
configuration on a cubic lattice \cite{kt}.
By construction these
are closed, orientable surfaces composed of plaquettes of the dual
lattice orthogonal to each frustrated link. In this way each plaquette
appears at most once in the construction of the surface $S$.
\vskip .3 cm
\centerline{\bf Tab. I}
\noindent{\small Graphs for the microscopic reconstruction of the
interface. Each cube is dual to a vertex $V_q$ with  $q$ edges of the
interface  as drawn in fig. 1. The  encircled sites have a spin
different from the other sites. Graphs with  different vertex
decompositions correspond to end-points of contact lines. The first
decomposition is for {\sl positive} vertices, while the other is for
{\sl negative} vertices.}
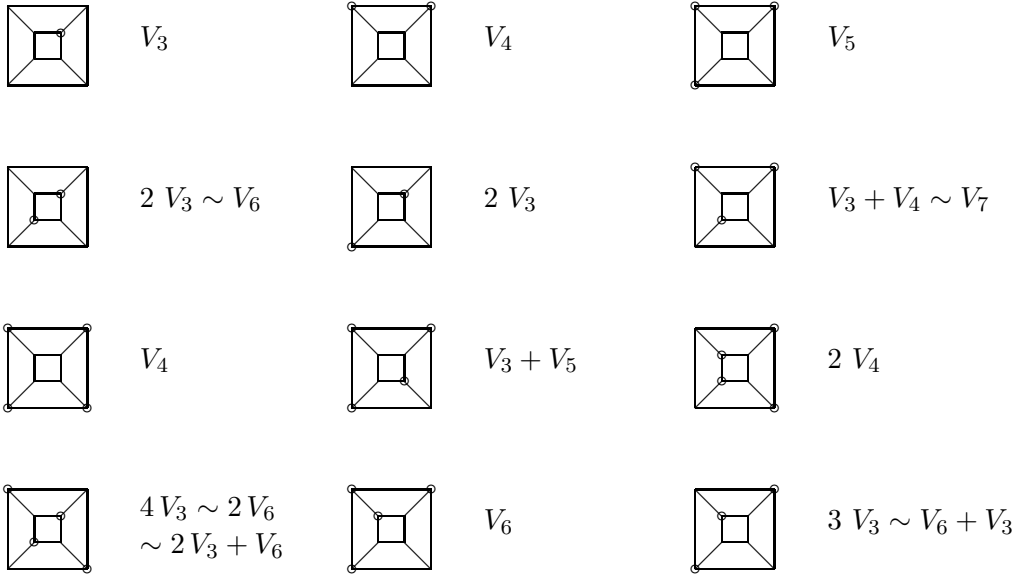
\begin{center}

\begin{picture}(380,60)(0,0)


\put(0,30){\line(1,0){30}}
\put(0,60){\line(1,0){30}}
\put(10,40){\line(1,0){10}}
\put(10,50){\line(1,0){10}}

\put(0,30){\line(0,1){30}}
\put(30,30){\line(0,1){30}}
\put(10,40){\line(0,1){10}}
\put(20,40){\line(0,1){10}}

\put(0,30){\line(1,1){10}}
\put(0,60){\line(1,-1){10}}
\put(20,50){\line(1,1){10}}
\put(20,40){\line(1,-1){10}}

\put(20,50){\circle{3}}

\put(50,45){$V_3$}


\put(130,30){\line(1,0){30}}
\put(130,60){\line(1,0){30}}
\put(140,40){\line(1,0){10}}
\put(140,50){\line(1,0){10}}

\put(130,30){\line(0,1){30}}
\put(160,30){\line(0,1){30}}
\put(140,40){\line(0,1){10}}
\put(150,40){\line(0,1){10}}

\put(130,30){\line(1,1){10}}
\put(130,60){\line(1,-1){10}}
\put(150,50){\line(1,1){10}}
\put(150,40){\line(1,-1){10}}

\put(130,60){\circle{3}}
\put(160,60){\circle{3}}

\put(180,45){$V_4$}


\put(260,30){\line(1,0){30}}
\put(260,60){\line(1,0){30}}
\put(270,40){\line(1,0){10}}
\put(270,50){\line(1,0){10}}

\put(260,30){\line(0,1){30}}
\put(290,30){\line(0,1){30}}
\put(270,40){\line(0,1){10}}
\put(280,40){\line(0,1){10}}

\put(260,30){\line(1,1){10}}
\put(260,60){\line(1,-1){10}}
\put(280,50){\line(1,1){10}}
\put(280,40){\line(1,-1){10}}

\put(260,60){\circle{3}}
\put(260,30){\circle{3}}
\put(290,60){\circle{3}}

\put(310,45){$V_5$}

\end{picture}

\begin{picture}(380,60)(0,0)


\put(0,30){\line(1,0){30}}
\put(0,60){\line(1,0){30}}
\put(10,40){\line(1,0){10}}
\put(10,50){\line(1,0){10}}

\put(0,30){\line(0,1){30}}
\put(30,30){\line(0,1){30}}
\put(10,40){\line(0,1){10}}
\put(20,40){\line(0,1){10}}

\put(0,30){\line(1,1){10}}
\put(0,60){\line(1,-1){10}}
\put(20,50){\line(1,1){10}}
\put(20,40){\line(1,-1){10}}

\put(20,50){\circle{3}}
\put(10,40){\circle{3}}

\put(50,45){$2~V_3\sim V_6$}


\put(130,30){\line(1,0){30}}
\put(130,60){\line(1,0){30}}
\put(140,40){\line(1,0){10}}
\put(140,50){\line(1,0){10}}

\put(130,30){\line(0,1){30}}
\put(160,30){\line(0,1){30}}
\put(140,40){\line(0,1){10}}
\put(150,40){\line(0,1){10}}

\put(130,30){\line(1,1){10}}
\put(130,60){\line(1,-1){10}}
\put(150,50){\line(1,1){10}}
\put(150,40){\line(1,-1){10}}

\put(130,30){\circle{3}}
\put(150,50){\circle{3}}

\put(180,45){$2~V_3$}


\put(260,30){\line(1,0){30}}
\put(260,60){\line(1,0){30}}
\put(270,40){\line(1,0){10}}
\put(270,50){\line(1,0){10}}

\put(260,30){\line(0,1){30}}
\put(290,30){\line(0,1){30}}
\put(270,40){\line(0,1){10}}
\put(280,40){\line(0,1){10}}

\put(260,30){\line(1,1){10}}
\put(260,60){\line(1,-1){10}}
\put(280,50){\line(1,1){10}}
\put(280,40){\line(1,-1){10}}

\put(260,60){\circle{3}}
\put(270,40){\circle{3}}
\put(290,60){\circle{3}}

\put(310,45){$V_3+V_4\sim V_7$}

\end{picture}
\begin{picture}(380,60)(0,0)


\put(0,30){\line(1,0){30}}
\put(0,60){\line(1,0){30}}
\put(10,40){\line(1,0){10}}
\put(10,50){\line(1,0){10}}

\put(0,30){\line(0,1){30}}
\put(30,30){\line(0,1){30}}
\put(10,40){\line(0,1){10}}
\put(20,40){\line(0,1){10}}

\put(0,30){\line(1,1){10}}
\put(0,60){\line(1,-1){10}}
\put(20,50){\line(1,1){10}}
\put(20,40){\line(1,-1){10}}

\put(0,30){\circle{3}}
\put(0,60){\circle{3}}
\put(30,30){\circle{3}}
\put(30,60){\circle{3}}

\put(50,45){$V_4$}


\put(130,30){\line(1,0){30}}
\put(130,60){\line(1,0){30}}
\put(140,40){\line(1,0){10}}
\put(140,50){\line(1,0){10}}

\put(130,30){\line(0,1){30}}
\put(160,30){\line(0,1){30}}
\put(140,40){\line(0,1){10}}
\put(150,40){\line(0,1){10}}

\put(130,30){\line(1,1){10}}
\put(130,60){\line(1,-1){10}}
\put(150,50){\line(1,1){10}}
\put(150,40){\line(1,-1){10}}

\put(130,30){\circle{3}}
\put(130,60){\circle{3}}
\put(160,60){\circle{3}}
\put(150,40){\circle{3}}

\put(180,45){$V_3+V_5$}


\put(260,30){\line(1,0){30}}
\put(260,60){\line(1,0){30}}
\put(270,40){\line(1,0){10}}
\put(270,50){\line(1,0){10}}

\put(260,30){\line(0,1){30}}
\put(290,30){\line(0,1){30}}
\put(270,40){\line(0,1){10}}
\put(280,40){\line(0,1){10}}

\put(260,30){\line(1,1){10}}
\put(260,60){\line(1,-1){10}}
\put(280,50){\line(1,1){10}}
\put(280,40){\line(1,-1){10}}

\put(270,40){\circle{3}}
\put(270,50){\circle{3}}
\put(290,30){\circle{3}}
\put(290,60){\circle{3}}

\put(310,45){$2~V_4$}

\end{picture}

\begin{picture}(380,60)(0,0)


\put(0,30){\line(1,0){30}}
\put(0,60){\line(1,0){30}}
\put(10,40){\line(1,0){10}}
\put(10,50){\line(1,0){10}}

\put(0,30){\line(0,1){30}}
\put(30,30){\line(0,1){30}}
\put(10,40){\line(0,1){10}}
\put(20,40){\line(0,1){10}}

\put(0,30){\line(1,1){10}}
\put(0,60){\line(1,-1){10}}
\put(20,50){\line(1,1){10}}
\put(20,40){\line(1,-1){10}}

\put(0,60){\circle{3}}
\put(10,40){\circle{3}}
\put(20,50){\circle{3}}
\put(30,30){\circle{3}}

\put(50,50){$4\,V_3\sim2\,V_6$}
\put(50,36){$\sim2\,V_3+V_6$}


\put(130,30){\line(1,0){30}}
\put(130,60){\line(1,0){30}}
\put(140,40){\line(1,0){10}}
\put(140,50){\line(1,0){10}}

\put(130,30){\line(0,1){30}}
\put(160,30){\line(0,1){30}}
\put(140,40){\line(0,1){10}}
\put(150,40){\line(0,1){10}}

\put(130,30){\line(1,1){10}}
\put(130,60){\line(1,-1){10}}
\put(150,50){\line(1,1){10}}
\put(150,40){\line(1,-1){10}}

\put(130,30){\circle{3}}
\put(130,60){\circle{3}}
\put(140,50){\circle{3}}
\put(160,60){\circle{3}}

\put(180,45){$V_6$}


\put(260,30){\line(1,0){30}}
\put(260,60){\line(1,0){30}}
\put(270,40){\line(1,0){10}}
\put(270,50){\line(1,0){10}}

\put(260,30){\line(0,1){30}}
\put(290,30){\line(0,1){30}}
\put(270,40){\line(0,1){10}}
\put(280,40){\line(0,1){10}}

\put(260,30){\line(1,1){10}}
\put(260,60){\line(1,-1){10}}
\put(280,50){\line(1,1){10}}
\put(280,40){\line(1,-1){10}}

\put(260,30){\circle{3}}
\put(270,50){\circle{3}}
\put(290,60){\circle{3}}

\put(310,45){$3~V_3\sim V_6+V_3$}

\end{picture}

\end{center}
A link belonging to $S$ is said to be regular if it glues two plaquettes
of $S$.
The only singularities which may appear on these surfaces are
links glueing four distinct plaquettes of $S$. These singular links form
contact lines of the surface which are sometimes improperly called
self-intersections. These contact lines are the only potential sources
of ambiguities in the  topological reconstruction of the surface. We
shall see shortly how to remove them in a simple, consistent
way.
A vertex of the surface $S$ is dual to an elementary cube of the lattice
of the Ising configurations. According to the distribution of spins
inside the cube, there are twelve distinct vertices as listed in Table
I. Some of them include singular links and may be interpreted as the
coalescence of distinct, regular vertices. Only the end-points of
contact lines can be decomposed into two or more inequivalent ways: for
instance the first graph in the second row of  the table can be thought
either as the coalescence of two vertices with three edges, or as a
single vertex with six edges and the singular link is splitted
accordingly into two different ways as drawn in figure 1.
\begin{center}
\begin{picture}(380,120)(0,0)

\put(0,30){\line(1,0){60}}
\put(0,90){\line(1,0){60}}
\put(0,30){\line(0,1){60}}
\put(60,30){\line(0,1){60}}
\put(20,50){\line(1,0){20}}
\put(25,45){$~_b$}
\put(20,50){\line(0,1){20}}
\put(10,60){$~_f$}
\put(20,70){\line(1,0){20}}
\put(25,75){$~_e$}
\put(40,50){\line(0,1){20}}
\put(40,60){$~_c$}
\put(0,30){\line(1,1){20}}
\put(11,40){$~_a$}
\put(40,70){\line(1,1){20}}
\put(0,90){\line(1,-1){20}}
\put(40,50){\line(1,-1){20}}
\put(20,50){\circle{5}}
\put(40,70){\circle{5}}
\put(46,75){$~_d$}
\put(80,50){$\leftrightarrow$}

\put(100,30){\line(1,0){20}}
\put(110,40){\line(1,0){10}}
\put(130,40){\line(1,0){20}}
\put(110,60){\line(1,0){40}}
\put(150,50){\line(1,0){10}}
\put(120,30){\line(0,1){20}}
\put(110,40){\line(0,1){20}}
\put(130,40){\line(0,1){20}}
\put(150,40){\line(0,1){20}}
\put(140,70){\line(0,-1){10}}
\put(100,30){\line(1,1){10}}
\put(120,30){\line(1,1){10}}
\put(120,50){\line(1,1){20}}
\put(150,40){\line(1,1){10}}

\put(105,35){$~_a$}
\put(151,55){$~_d$}
\put(120,43){$~_b$}
\put(135,50){$~_c$}
\put(110,50){$~_f$}
\put(140,75){$~_e$}

\put(205,5){$~_a$}
\put(251,25){$~_d$}
\put(220,13){$~_b$}
\put(237,20){$~_c$}
\put(210,20){$~_f$}
\put(242,45){$~_e$}

\put(205,75){$~_a$}
\put(241,85){$~_d$}
\put(215,90){$~_f$}
\put(230,95){$~_e$}
\put(220,65){$~_b$}
\put(235,70){$~_c$}

\put(170,50){$\to$}

\put(200,70){\line(1,0){20}}
\put(210,80){\line(1,0){20}}
\put(230,80){\line(1,0){20}}
\put(210,100){\line(1,0){20}}
\put(240,90){\line(1,0){20}}
\put(220,70){\line(0,-1){20}}
\put(210,80){\line(0,1){20}}
\put(230,80){\line(0,1){20}}
\put(230,80){\line(0,-1){20}}
\put(250,80){\line(0,-1){20}}
\put(240,110){\line(0,-1){20}}
\put(200,70){\line(1,1){10}}
\put(230,80){\line(1,1){10}}
\put(220,70){\line(1,1){10}}
\put(230,100){\line(1,1){10}}
\put(250,80){\line(1,1){10}}
\put(230,60){\line(1,0){20}}
\put(220,50){\line(1,1){10}}

\put(200,0){\line(1,0){20}}
\put(210,10){\line(1,0){10}}
\put(232,10){\line(1,0){20}}
\put(210,30){\line(1,0){20}}
\put(232,30){\line(1,0){20}}
\put(252,20){\line(1,0){10}}
\put(220,0){\line(0,1){20}}
\put(210,10){\line(0,1){20}}
\put(230,10){\line(0,1){20}}
\put(232,10){\line(0,1){20}}
\put(252,10){\line(0,1){20}}
\put(242,40){\line(0,-1){10}}
\put(200,0){\line(1,1){10}}
\put(220,0){\line(1,1){10}}
\put(220,20){\line(1,1){10}}
\put(232,30){\line(1,1){10}}
\put(252,10){\line(1,1){10}}
\end{picture}
\end{center}
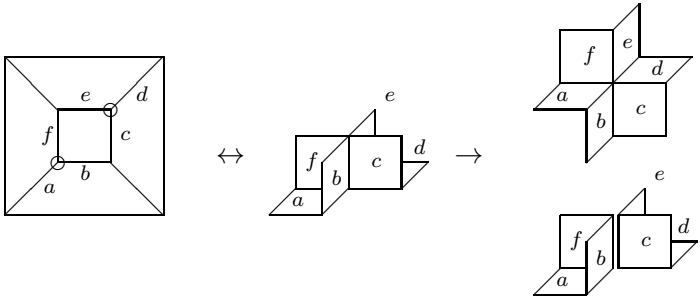
\centerline{\bf Fig. 1}
{\small Example of decodification of a graph of Table I.
This cube has six frustrated links, labelled by $a,b,\dots,f$, which
are dual to six plaquettes forming a vertex with a singular link.
 It can be considered either as a 6-vertex, or as two 3-vertices. This
choice depends on the sign of the magnetization of the cube, which can
be in this case $\pm 4$.}
\vskip .2 cm
Choosing arbitrarily this splitting procedure for each end-point may
 generate  global obstructions in the surface
reconstruction, due to a constraint which links together the splitting
of two end-points connected by a contact line. In order to formulate
this constraint, it is useful to attribute a sign to each singular vertex
according to the sign of the magnetization of the corresponding
elementary cube. Then, working out some explicit example, it is easy to
convince oneself that there are no global obstructions if one chooses
different splittings for vertices of different sign. More precisely, the
only constraint to be fulfilled in the replacement of each singular
link with a pair of regular ones is that two end-points connected by a
contact line must be  splitted in the same way if they have the same
sign, or in the opposite way if their sign is different. The two
decompositions appearing for singular vertices in table I correspond
precisely to the two possible signs of the vertex. The only case in
which the surface reconstruction is not immediate is the first graph of
the last row of tab. I, corresponding to a vertex with six singular
lines: the three possible decompositions cannot be selected by the sign
of its magnetization (which is zero), but by the signs of the end-points
of the six contact lines.
\vskip .3 cm
The main consequence of the above construction is that we have obtained
a simple rule to assign unambiguously to each Ising configuration
a set of self-avoiding closed random surfaces. Notice that this fact
agrees with a result of David \cite{fd} who found that a gas of
self-avoiding surfaces in a special three-dimensional lattice belongs
to the same universality class of the Ising model.
\vskip .3cm
We may now evaluate the genus $h$ of each self-avoiding surface through the
Euler relation
\eq
F-E+V=\chi(S)=2-2h
\label{euler}
\en
where $F$ is the number of faces (plaquettes), $E$ the number of
edges, $V$ the number of vertices and $\chi$ the Euler characteristic;
the genus $h$ gives the number of handles. Denoting by $N_q$ the number
of vertices of coordination number $q$, we have obviously, according to
table I,
\eq
V=N_3+N_4+N_5+N_6+N_7~~;~4F=2E=3N_3+4N_4+5N_5+6N_6+7N_7~~,
\en
which gives at once an even simpler expression for the genus of this
kind of surfaces
\eq
4(V-F)=N_3-N_5-2N_6-3N_7=8-8h~~.
\label{VF}
\en
\vskip .3 cm
{\bf 3. Results}
\vskip .3cm

We are interested on the topology of the interface between two
macroscopic domains of opposite magnetization. For this reason we
consider very elongated lattices with periodic boundary conditions in
the two short directions (denoted in the following by $L$) and
{\sl antiperiodic} boundary conditions in the long direction (denoted by
 $L_z$). This forces the formation of an odd number of interfaces in the
$L_z$ direction. We then isolate one of these interfaces by
reconstructing all the spin clusters of the configuration, keeping the
largest one and flipping the others. Note, as a side remark, that for
the values of $\beta$, $L$ and $L_z$ we have studied, typical
configurations contained only one macroscopic cluster,
besides a huge number of microscopic ones. It is
now easy to evaluate  area of the interface by counting simply the
number of frustrated links of this cleaned configuration.
\vskip .3 cm
The growth of this area as a function of the size of the lattice gives
a first description of the fractal behaviour of the interface:
in a set of Monte Carlo simulations near the critical
point on elongated lattices $L^2\times L_z$ with $L_z\ge120$ and
$8\le L\le16$ (see table II) we found that the mean area $<F>$ of the
interface is a strongly varying function of the transverse size $L$,
well parametrized by the following power law
\eq
<F>=\kappa\,L^{d_H}~~,
\label{dh}
\en
with $d_H\sim3.7$ and $\kappa\sim0.47$~. As a  consequence,
 a sizeable fraction of the lattice is invaded by the
interface, and a small increasing of the transverse lattice section
implies a rapid growth of the area of the interface, so we have to take
very elongated lattices in order to avoid wrapping of the interface
around the antiperiodic direction, which would give rise to unwanted
finite volume effects.
\vskip .3 cm
\def\e{\rm e}
Each interface $S$ of area $F$ of an arbitrary Ising configuration
contributes to the partition function  simply with a term
$\e^{-\beta F}$; then we can define the following generating functional
$Z$
\eq
Z(\beta)=\sum_{F}\sum_{\#~{\rm surfaces~of~area~F}}\,\e^{-\beta F}=
\sum_F\sum_h Z_h(F)~~~,
\en
where $Z_h(F)$ is the partition function for a surface of area $F$ and
genus $h$. Since we are dealing with  macroscopic surfaces, we can
consistently  assume that their multiplicity is  well described by the
asymptotic behaviour of the entropy of large random surfaces \cite{adf},
which yields
\eq
Z_h(F)_{F\to \infty}\sim \; F^{b_\chi-1}\e^{\mu_c F}\e^{-\beta F}=
F^{b_\chi-1}\e^{-\mu F}~~~,
\label{zhf}
\en
where, using the terminology of two-dimensional quantum gravity (2DQG),
$\mu$ is the cosmological constant and the exponent $b_\chi$ is a
function of the Euler characteristic $\chi$. In the continuum theory
of 2DQG this exponent can be evaluated exactly for the coupling to
conformal matter of central charge $c\leq1$ \cite{ddk}. The result
is
\eq
b_\chi=-\frac{b}{2}\chi=b(h-1)~~~,
\label{bc}
\en
with
\eq
b=\frac{25-c+\sqrt{(1-c)(25-c)}}{12}~~~.
\label{b}
\en
$b$ is a monotonically decreasing function of $c$ for $c\leq 1$
and becomes complex for $c > 1$, where the surface get crumpled and
these formulas lose any physical meaning. It is known that for $c=1$
there are logarithmic corrections to eq.({\ref{zhf}), while for $c>1$ it
is only known that the number of surfaces of given area  is
exponentially bounded \cite{adf}. It turns out that eq.({\ref{zhf}) fits
well to our numerical data.

We can evaluate the exponent $b_\chi$ for the interface by measuring the
mean area at fixed genus. Indeed defining
\eq
Z_h(\mu)=\sum_F Z_h(F)~~~,
\en
we have
\eq
<F>_h=-\frac{\partial}{\partial\mu}\log Z_h(\mu)\sim
\frac{b_\chi}{\mu}~~~.
\label{fh}
\en
In figure 2 we report a typical outcome of this analysis. Actually
$<F>_h$ is dominated by a linear term, like in eq.(\ref{bc}),
but there is also a small, negative quadratic contribution
which  probably accounts for the self-avoidness constraint: when
the number of handles $h$ is very large the excluded volume effects
(proportional to $h^2$) become important. The data are well fitted by
\eq
<F>_h=\frac{a}{\mu}+\frac{b}{\mu}h+\frac{d}{\mu}h^2~~.
\label{fit}
\en
The slope $b$ can be called the entropy exponent, and $a$ is related
to the strings susceptibility by $a=\gamma_s-2$.
Comparing eq.(\ref{fit}) with eq.(\ref{bc}) we have of course, for $c\leq1$,
$d=0$
and
\eq
b=-a\equiv2-\gamma_s~~.
\label{ee}
\en
Since we are dealing with surfaces of very high genus $h\sim10^2$, we
can evaluate $b/\mu$ and $d/\mu\sim-10^{-4}b/\mu$ very accurately,
while the constant $a$ is affected by large errors  and cannot used to
evaluate $\gamma_s$, nevertheless we shall argue shortly that for our
surfaces eq.(\ref{ee}) is no longer true.

Combining eq.(\ref{fh}) with other physical quantities at fixed genus
it is possible to evaluate the cosmological constant $\mu$ as a function
of $\beta$ and of the lattice size. In particular, using the
derivative $\frac{\partial Z_h(\mu)}{\partial h}$, we get easily the
following equation
\eq
\log(1/\mu)=-\psi(b_\chi)+<\log F>~~~,
\label{mu}
\en
where $\psi$ denotes the logarithmic derivative of the Euler $\Gamma$
function. This formula fits well the numerical data for large areas
and we used it  to evaluate $\mu$.
The results are reported in table II as well as the entropy exponent
$b$ defined in eq.(\ref{fit})~.
It turns out that  $b$ is affected by rather strong finite size effects,
being a  decreasing function of the lattice size $L$ an hence of the
area. Similar finite size effects have been observed for the string
susceptibility exponent in the planar quantum gravity coupled to
$c\geq1$ matter \cite{ajt}. Note however that, if one assumes
eq.(\ref{ee}), one should conclude that $\gamma_s$ exceeds the
theoretical upper bound $\gamma_s=\frac{1}{2}$ \cite{adfo}.
A possible way out is that eq.(\ref{ee}) is not fulfilled by this kind
of surfaces, indeed the argument leading to such upper bound assumes
$h=0$ and gives no restriction on $b$. On the other hand there is a
renormalization group argument \cite{bz} showing clearly that
eq.(\ref{ee}) is  justified only for $c\leq1$.
\vskip.6 cm
\centerline {\bf Tab. II}
{\it Mean area $<F>$, entropy exponent $b$, cosmological constant $\mu$
 and renormalized cosmological constant $\mu_R$ of the interface in a
set of elongated lattices of shape $L^2\times L_z$  at
$\beta-\beta_c=0.0059$~.}
$$\vbox {\offinterlineskip
\halign  { \strut#& \vrule# \tabskip=.5cm plus1cm
& \hfil#\hfil & \vrule# & \hfil# \hfil
& \vrule# & \hfil# \hfil
& \vrule# & \hfil# \hfil &
& \vrule# & \hfil# \hfil &\vrule# \tabskip=0pt \cr \noalign {\hrule}

&& $L^2\times L_z$ && $<F>$  && $b$  && $\mu$
&& $\mu_R$ & \cr \noalign {\hrule}
&& $8^2\times 120$ && $1057(10)$ && $0.73(20)$ &&
$0.0141(39)$ && $0.00122(4)$ & \cr \noalign {\hrule}
&& $10^2\times120$ && $2434(19)$ && $0.61(15)$ &&
$0.0097(24)$ && $0.000613(15)$ & \cr \noalign {\hrule}
&& $12^2\times180$ && $4719(47)$ && $0.30(3)$ &&
$0.0047(5)$ && $0.000347(4)$ & \cr \noalign {\hrule}
&& $14^2\times240$ && $8506(103)$ && $0.19(3)$ &&
$0.0026(5)$ && $0.000208(3)$ & \cr \noalign {\hrule}
&& $16^2\times320$ && $14803(249)$ && $0.11(2)$ &&
$0.0015(2)$ && $0.000125(5)$ & \cr \noalign {\hrule}
}}$$

It is also possible to define a renormalized cosmological constant
$\mu_R$ by considering the sum over all the genera $Z(F)=\sum_h Z_h(F)
$~. It turns out that for large areas the distribution of surfaces is
accurately described by  an exponential fall off of the type
\eq
Z(F)_{F\to\infty}\sim c\e^{-\mu_R F}~~~,
\label{mur}
\en
as shown in fig.3. Comparing this equation with eq.(\ref{zhf}), we
argue that the sum over all the topologies produces as  net effect
an effective interface which behaves as a smooth surface with a
different cosmological constant. Actually the renormalization effect is
very large: $\mu_R$ is about one order of magnitude smaller than the
unrenormalized quantity $\mu~$  (see table II). This phenomenon seems
the two-dimensional analogue of a quantum gravity effect described by
Coleman  \cite{sc}, who argued that the sum over topologies has the
effect of making the cosmological constant vanishing small.

In conclusion, we have found a  simple way to generate high-genus
self-avoiding random surfaces. The most interesting  question is
of course whether their study will lead to new insights in the physics
of random surfaces.
\vskip .3 cm
{\bf Acknowledgements.}
We thank A.Pelissetto and A. Sokal for sending us their very efficient
program of cluster reconstruction, and D. Boulatov, V. Kazakov and A.
Migdal for enlightening discussions.

\hrule
\vskip .6 cm
\centerline{\bf Figure Captions}
\begin{description}
\item{\sl Figure 2}
The mean area of the interface as a function of the genus in a
lattice of size $12^2\times240$ at $\beta-\beta_c=0.0044$~.
\item {\sl Figure 3}
The multiplicity of interface configurations summed over all the
topologies as a function of the area in a lattice of size
$14^2\times240$ at $\beta-\beta_c=0.0059$. The lower set of data is the
multipicity a fixed genus $h=70$. The stright line is the fit to
eq.(\ref{mur}).
\end{description}}}
\end{document}